\documentclass[12pt,preprint]{aastex}

\def\vec#1{\mbox{\boldmath $#1$}}

\slugcomment{Not to appear in Nonlearned J., 45.}

\shorttitle{3D Magnetic Field and Its Stability of AR12192}
\shortauthors{S./ Inoue.}

\begin{document}


\title{Structure and Stability of Magnetic Fields in Solar Active Region
        12192 Based on Nonlinear Force-Free Field Modeling}


\author{S.\ Inoue}
\affil{Max-Planck-Institute for Solar System Research, Justus-von-Liebig-Weg 
       3 37077 G\"{o}ttingen Germany}
\altaffiltext{1}{Institute for Space-Earth Environment Research, Nagoya University \\
       Furo-cho, Chikusa-ku, Nagoya, 464-8601, Japan}
\altaffilmark{1}
\email{inoue@mps.mpg.de}

\author{K.\ Hayashi}
\affil{Institute for Space-Earth Environment Research, Nagoya University, Chikusaku 
       Furo-Cho Nagoya 464-8601 Japan}           
           
\author{K.\ Kusano}
\affil{Institute for Space-Earth Environment Research, Nagoya University, Chikusaku 
       Furo-Cho Nagoya 464-8601 Japan}    
\altaffiltext{2}{Japan Agency for Marine-Earth Science and Technology(JAMSTEC),
       Kanazawa-ku, Yokohama, Kanagawa,236-0001, Japan}    
    
  \begin{abstract}
   We analyze a three-dimensional (3D) magnetic structure and its stability in large solar 
   active region(AR) 12192, using the 3D coronal magnetic field constructed under a 
   nonlinear force-free field (NLFFF) approximation. In particular, we focus on the magnetic 
   structure that produced an X3.1-class flare which is one of the X-class flares observed in 
   AR 12192. According to our analysis, the AR contains multiple-flux-tube system, {\it e.g.}, 
   a large flux tube, both of whose footpoints are anchored to the large bipole field, under 
   which other tubes exist close to a polarity inversion line (PIL). These various flux tubes of 
   different sizes and shapes coexist there. In particular, the later are embedded along the PIL, 
   which produces a favorable shape for the tether-cutting reconnection and is related to the 
   X-class solar flare. We further found that most of magnetic twists are not released even 
   after the flare, which is consistent with the fact that no observational evidence for major 
   eruptions was found. On the other hand, the upper part of the flux tube is beyond a critical 
   decay index, essential for the excitation of torus instability before the flare, even though 
   no coronal mass ejections (CMEs) were observed. We discuss the  stability of the complicated 
   flux tube system and suggest  the reason for the existence of the stable flux tube. In addition, 
   we further point out a possibility for tracing the shape of flare ribbons, on the basis of a detailed 
   structural analysis of the NLFFF before a flare.   
  \end{abstract}

    \section{Introduction}
     The solar storms associated with solar flares, coronal mass ejections (CMEs) 
    strongly affect the electromagnetic environment in our geospace
    (\citealt{2014SpWea..12..380K}). Therefore, understanding their onset and 
    dynamics is very important not only for obtaining new knowledge of the 
    nonlinear dynamics of the solar coronal plasma but also for establishing space 
    weather forecasting.  
  
\begin{figure}
  \epsscale{1.}
  \plotone{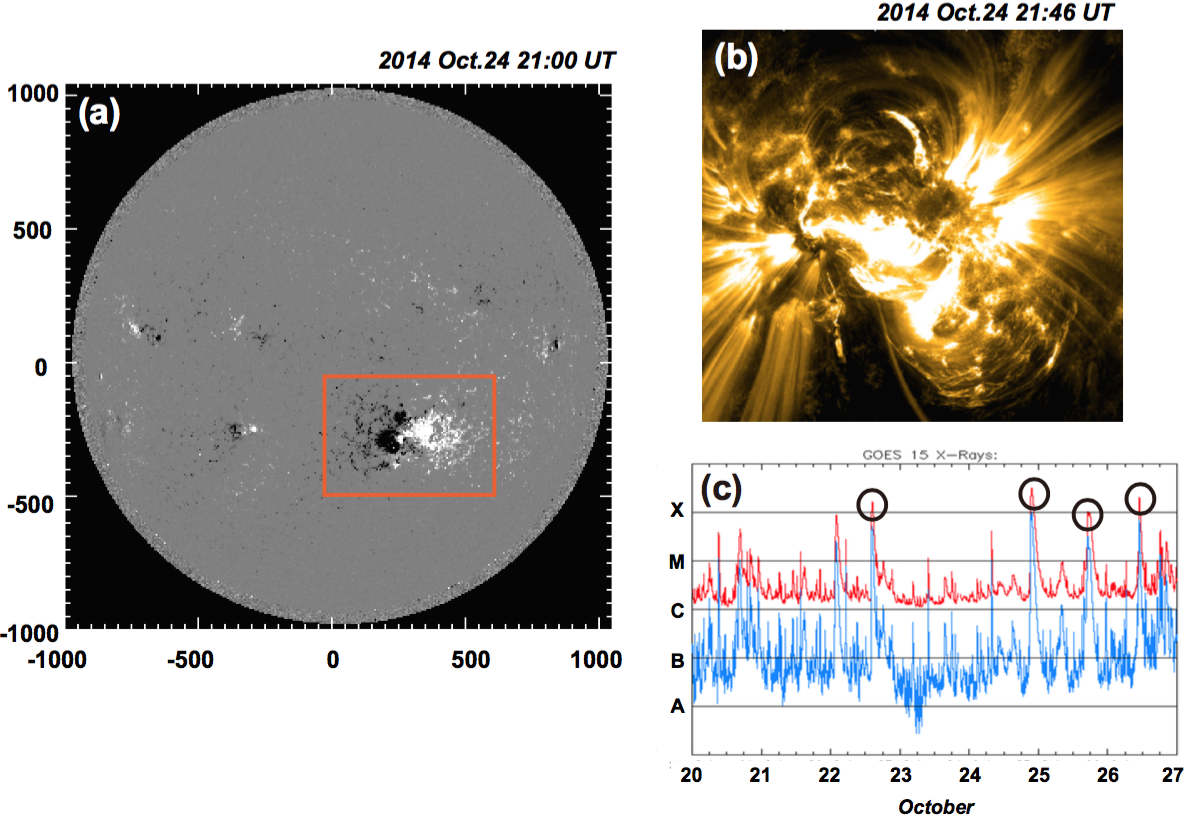}
  \caption{
               (a) Full-disk image of a line-of-sight component of the solar magnetic 
                     field observed at 21:00 UT on October 24 in an onset phase of an 
                     X3.1-class flare by {\it SDO}/HMI.
               (b) EUV image during the X3.1-class flare observed at 21:46 UT by
                    {\it SDO}/AIA 171 \AA.
               (c) Time profile of the X-ray flux measured by the GOES 12
                    satellite from October 20 to 27, 2014. The solar X-ray outputs 
                    in the 1$-$8 \AA \ and  $0.5-4.0$ \AA \ passband are plotted, 
                    respectively. Four X-class flares  were observed during 
                    this period.}
  \label{f1}
  \end{figure}

      Solar active region (AR) 12192, which is the largest AR to date  in solar cycle 
    24 appeared in the middle of October 2014, and is shown in Figure \ref{f1}(a). 
    Interestingly, this AR produced four X-class flares, (one of which is shown in 
    Figure \ref{f1}(b)), in just a week, October 20-27, 2014, as shown in Figure \ref{f1}(c). 
    Nevertheless, CMEs were not observed to accompany these huge flares, whereas 
    \cite{2006ApJ...650L.143Y} reported that flares of such a magnitude are highly 
    likely to be accompanied by CMEs. The ejected flux rope formed via the flare process, 
    which would be central to the occurrence of a CME, is  considered to be inhibited 
    by the surroundings, a phenomenon that is well known and often observed as 
    confined flares or eruptions, which have been well studied. For instance, 
    \cite{2003ApJ...595L.135J} observed that a filament driven by kink instability was 
    suddenly launched, but it stopped during its ascent, and consequently could not exhibit 
    an eruption. \cite{2010ApJ...725L..38G} also reported a good example of confined 
    eruption of a filament on 2005 May 27 in AR 10767, even though the filament 
    showed a strong writhing motion due to highly twisted lines satisfying the 
    condition for the kink instability.  More recently, \cite{2014ApJ...795....4J} analyzed the 
    confined eruption associated with a C-class flare on 2014 January 1 in AR11938. 
    These dynamics showed that two filaments merged by tether-cutting reconnection 
    (\citealt{2001ApJ...552..833M}), consequently, a long helical filament was formed, and 
    plasma flow was observed as a reconnection jet with a velocity of more than 200(km/s) 
    along a new long filament. However, no eruptions were seen. \cite{2014ApJ...785...88Z} 
    also reported no filament eruptions associated with an M6.0 flare on August 3, 2011 in AR 
    11261, although a filament eruption occurred later, on August 4. 

  They reported that the configuration is stable against the torus instability
  (\citealt{2006PhRvL..96y5002K}). The torus instability is induced by collapse of the force 
  balance between the upward force on a flux tube, {\it i.e.}, the hoop force driven by a current, 
  and the downward force. The height at which the loop becomes unstable to the torus mode is 
  determined  by the decay index in the external field derived from the flux tube's surroundings.
  (\citealt{2007AN....328..743T},
   \citealt{2010ApJ...718.1388D}),
  and is specified in the AR of interest, {\it e.g.},
  \cite{2010ApJ...725L..38G},
  \cite{2014ApJ...795....4J},
  \cite{2014ApJ...785...88Z},
  \cite{2015ApJ...804L..28S}.
  Therefore, stability analysis, in particular, analysis of relationship between the 
  height at which torus instability is excited and the height of the flux tube, is 
  essential for understanding the onset of eruptions. To perform such analysis, 
  we need to determine the magnetic field structures in detail. On the other hand, 
  we cannot directly measure the magnetic field in the solar corona, although 
  extreme ultraviolet (EUV) or X-ray images help us to indirectly determine the 
  flux tube height ({\it e.g.}, \citealt{2014ApJ...795....4J}).

  Despite the lack of methods for directly determining the magnetic field in an AR, 
  nonlinear force-free field (NLFFF) extrapolation(\citealt{2012LRSP....9....5W}) 
  is a robust tool for obtaining information on the three-dimensional(3D) coronal 
  magnetic field. The NLFFF can be extrapolated from three components of the 
  photospheric magnetic field which have been observed by state-of-the-art 
  instruments; the Solar Optical Telescope (SOT:\citealt{2008SoPh..249..167T}) on 
  board {\it Hinode} (\citealt{2007SoPh..243....3K}) and the Helioseismic and Magnetic 
  Imager (HMI:\citealt{2014SoPh..289.3483H}) on board {\it Solar Dynamics 
  Observatory (SDO)}(\citealt{2012SoPh..275..207S}). NLFFF approaches have been 
  used to successfully reconstruct twisted field lines
  ({\it e.g.},
   \citealt{2009ApJ...693L..27C},
   \citealt{2010ApJ...725L..38G},
   \citealt{2014ApJ...780...55J},
   \citealt{2014ApJ...788..182I},
   \citealt{2014Natur.514..465A},
   \citealt{2015ApJ...788..182I}),
  some of these studies analyzed the stability analysis against an ideal 
  magnetohydrodynamic (MHD) instability and tried to explain the onset 
  and dynamics of solar eruptions. Regarding AR12192, \cite{2015ApJ...804L..28S} 
  already showed the configuration of the 3D magnetic field but did not address 
  the MHD instability in detail. AR 12192 showed complicated magnetic structure, 
  implying that various flux tubes of different sizes and shapes could coexist there. 
  Therefore, stability analysis should be treated carefully.

   In this study, we first extrapolate the 3D magnetic field of AR12192 under the 
  NLFFF approximation(\citealt{2014ApJ...780..101I}), in order to clarify its 3D 
  complex magnetic structure. In particular, we detect the flux tubes
  making up a knotty magnetic field and discuss the stability. Through 
  these analyses,  we eventually answer the question of why no solar eruption was 
  occurred even though the X3.1-class flare was clearly observed. The rest 
  of this paper is constructed as follows. The observations and numerical method 
  are described in section 2. The results and discussion are presented in sections 3 
  and section 4, respectively. Our conclusions are presented in section 5. 

  \section{Observation and Numerical Method}  
  \subsection{Observation}
  The vector magnetic fields in the SHARP format 
  (\citealt{2014SoPh..289.3549B}) are used as the boundary condition of 
  our NLFFF calculation, which is available online
  \footnote{Data are available at http://jsoc.stanford.edu/new/HMI/HARPS.html}.
  These data are given in a cylindrical equi-area projection. We choose observations 
  at 19:00 UT and 23:00 UT on October 24, approximately 2 h before and after the 
  X3.1-class flare, by HMI/{\it SDO} (see Figure \ref{f2}(a)). The data cover a 324 
  $\times$ 270 (Mm$^2$) region, that is divided into a 900 $\times$ 750 grid and 
  includes a large bipole shaped sunspot in AR12192. They were obtained using 
  the very fast inversion of the Stokes vector algorithm (\citealt{2011SoPh..273..267B}) 
  based on the Milne$-$Eddington approximation. The minimum energy method 
  (\citealt{1994SoPh..155..235M};\citealt{2009SoPh..260...83L}) was used to resolve 
  the 180$^\circ $ ambiguity in the azimuth angle of the magnetic field. In this study, 
  the vector field is preprocessed in accordance with \cite{2006SoPh..233..215W}. 
  We also used an EUV image of the  coronal magnetic fields taken by the Atmospheric 
  Imaging Assembly (AIA; \citealt{2012SoPh..275...17L}) on board {\it SDO}, for comparison 
  with the NLFFF.
  
  \subsection{NLFFF Extrapolation}
  The NLFFF extrapolation follows the MHD relaxation method developed 
  by \cite{2014ApJ...780..101I}. The considered equations and parameters are identical 
  as  adapted in \cite{2014ApJ...788..182I} and \cite{2015ApJ...788..182I}.  
  The basic equations are:
      
    \begin{equation}
    \rho=|\vec{B}|,
    \label{rho_eq}
    \end{equation}  
      
  \begin{equation}
  \frac{\partial \vec{v}}{\partial t} 
                        = - (\vec{v}\cdot\vec{\nabla})\vec{v}
                          + \frac{1}{\rho} \vec{J}\times\vec{B}
                          + \nu\vec{\nabla}^{2}\vec{v},
  \label{mhd1_eq}
  \end{equation}

  \begin{equation}
  \frac{\partial \vec{B}}{\partial t} 
                        =  \vec{\nabla}\times(\vec{v}\times\vec{B}
                        -  \eta\vec{J})
                        -  \vec{\nabla}\phi, 
  \label{induc_eq}
  \end{equation}

  \begin{equation}
  \vec{J} = \vec{\nabla}\times\vec{B},
  \end{equation}
  
  \begin{equation}
  \frac{\partial \phi}{\partial t} + c^2_{h}\vec{\nabla}\cdot\vec{B} 
    = -\frac{c^2_{h}}{c^2_{p}}\phi,
  \label{div_eq}
  \end{equation}
  where $\rho$, $\vec{B}$,   $\vec{v}$, $\vec{J}$, $\phi$ are the 
  pseudo density, magnetic flux density, the velocity, the electric current density, 
  and the convenient potential, respectively. The pseudo density is 
  assumed to be proportional to $|\vec{B}|$ in order make the Alfven 
  speed relax in space(\citealt{1996ApJ...466L..39A}). 
   The scalar $\phi$ was introduced by 
  \cite{2002JCoPh.175..645D} to minimize any deviations from 
  $\nabla\cdot\vec{B}=0$. The length, magnetic field, density, velocity, time 
  and electric current density are normalized by
  $L^{*}$ = 324 Mm,  
  $B^{*}$ = 2800 G, 
  $\rho^{*}$ = $|B^{*}|$,
  $V_{A}^{*}\equiv B^{*}/(\mu_{0}\rho^{*})^{1/2}$,    
  where $\mu_0$ is the magnetic permeability,
  $\tau_{A}^{*}\equiv L^{*}/V_{A}^{*}$, and     
  $J^{*}=B^{*}/\mu_{0} L^{*}$,    
  respectively.  The non-dimensional viscosity $\nu$ is set as a constant  
  $(1.0\times 10^{-3})$, and the non-dimensional resistivity $\eta$ is expressed 
  by a functional relation as
  \begin{equation}
  \eta = \eta_0 + \eta_1 \frac{|\vec{J}\times\vec{B}||\vec{v}|}{\vec{|B|}},
  \end{equation} 
  where $\eta_0 = 5.0\times 10^{-5}$ and $\eta_1=1.0\times 10^{-3}$ in 
  the non-dimensional unit. The second term is introduced to accelerate 
  the relaxation to the force free field, particularly in weak field region. 
  The other parameters $c_h^2$ and $c_p^2$ are fixed to constants 0.04 and 0.1, 
  respectively.  The velocity is controlled as follows. If the value of $v^{*}$ 
  becomes larger than $v_{max}$(here set to 0.01), then the velocity 
  is modified as follows: $\vec{v} \Rightarrow (v_{max}/ v^{*}) \vec{v}$. This process 
  can suppress large discontinuities produced between the bottom and inner domain.
  
  The potential field is given as the initial condition, which is extrapolated only 
  from the normal component of the photospheric magnetic field. During the 
  iteration, three components of the magnetic field are fixed at each boundary while 
  the velocity is fixed to zero and von Neumann condition $\partial /\partial n$=0 
  is imposed on $\phi$. Note that the bottom boundary is fixed according to 
   \[
    \vec{B}_{bc} = \zeta \vec{B}_{obs} + (1-\zeta) \vec{B}_{pot},
   \]
  where $\vec{B}_{bc}$  is the horizontal  component which is determined by 
  a linear combination of the observed magnetic field ($\vec{B}_{obs}$) and the 
  potential magnetic field ($\vec{B}_{pot}$). $\zeta$ is a coefficient ranging 
  from 0 to 1. When $R=\int |\vec{J}\times\vec{B}|^2$dV, which is calculated over the 
  interior of the computational domain, falls below a critical value denoted by $R_{min}$ 
  during the iteration, the value of the parameter $\zeta$ is increased to $\zeta = \zeta + d\zeta$. 
  In this paper, $R_{min}$ and d$\zeta$ have the values $5.0 \times 10^{-3}$ and 0.02, 
  respectively. If $\zeta$ becomes equal to 1, $\vec{B}_{bc}$ is completely consistent with 
  the observed data. This process also helps to avoid a sudden jump from the boundary into 
  the domain. A numerical box with dimensions of 324 $\times$ 270 $\times$ 270 (Mm$^3$) is 
  given as 1 $\times$ 0.8333 $\times$ 0.8333 in its non-dimensional value. A 3 $\times$ 3 
  binning of the original data yields a grid resolution of 300 $\times$ 250 $\times$ 250. 

  \section{Results}
  \subsection{Overview of 3D Magnetic Structure Before the X3.1-Class Flare}
\begin{figure}
  \epsscale{1.}
  \plotone{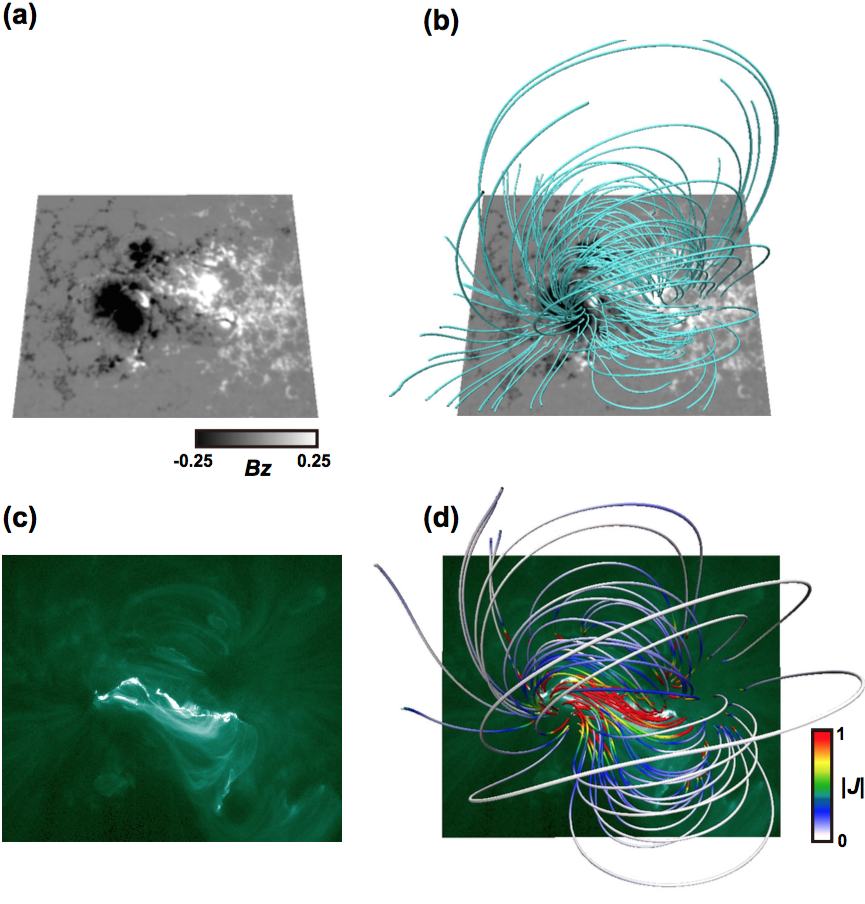}
  \caption{
       (a) Normal component $B_z$ of the photospheric magnetic field 
            observed at 19:00 UT on October  24, which is approximately 
            2h before the X3.1-class flare. 
       (b) Filed lines of the NLFFF are plotted over (a).
       (c) The EUV image observed at 21:10:50 UT on October 24 2014 
            by {\it SDO}/AIA.
       (d) Field lines are plotted over (c). Color indicates the strength of 
            the current density ($|\vec{J}|$). 
          }
  \label{f2}
  \end{figure}

   Figure \ref{f2}(a) shows the photospheric magnetic field obtained from {\it SDO}/HMI at 
  19:00 UT on October 24 2014,  which is approximately 2h before the X3.1-class 
  flare; the NLFFF calculation was applied to this field as the bottom boundary 
  condition. We can see large negative and positive sunspots in the center of the 
  field, whereas a region close to the PIL between them is filled with complex 
  magnetic structure of mixed positive and negative polarities. Figure \ref{f2}(b) 
  shows the 3D structure of the magnetic field lines of the NLFFF based on the 
  photospheric magnetic field shown in Figure \ref{f2}(a). We found the sheared 
  field lines between the major sunspots, whereas the area outside of them is filled 
  with the closed field lines in a potential field like structure. Figure \ref{f2}(c) shows 
  an EUV image obtained from {\it SDO}/AIA 94 \AA, whose covers a region identical 
  in size to that in Figure \ref{f1}(a). The field lines of the NLFFF are superimposed on 
  the EUV image in Figure \ref{f2}(d); the color indicates the strength of the current 
  density $|\vec{J}|=\sqrt{j_x^2 + j_y^2 + j_z^2}$. Field lines having strong current 
  density are obviously concentrated in the central region, in which the EUV is also 
  strongly enhanced, as shown in Figure \ref{f2}(c) 
  
  \subsection{Magnetic Twist of the NLFFF Before the X3.1 Class Flare}
   To understand the physical aspects of the NLFFF,  we have to detect the 
  essential feature in the complicated magnetic field in Figure \ref{f2}(b). In 
  particular, because a flux tube, which is a bundle of highly twisted lines, is 
  a major candidate for the driver of solar eruptions,  it plays an important 
  role in eruption onset and we need to correctly understand its stability. First, 
  to locate the flux tube, we calculate the magnetic twist, which is defined as 
  follows:
  \begin{equation}
   T_n=\frac{1}{4\pi}\int \frac{\vec{J}_{||}}{|\vec{B}|} dl,
  \label{twist}
  \end{equation}
  where $\vec{J}_{||}(=\vec{J}\cdot \vec{B}/|\vec{B}|)$ is the field-aligned 
  current. The magnetic twist is a turn of the field lines corresponding to the 
  magnetic helicity generated by the field-aligned current
  (\citealt{2006JPhA...39.8321B}), and has been applied to several ARs 
  ({\it e.g.}, \citealt{2011ApJ...738..161I}; \citealt{2012ApJ...760...17I};
   \citealt{2013ApJ...770...79I}).
  
\begin{figure}
  \epsscale{1.}
  \plotone{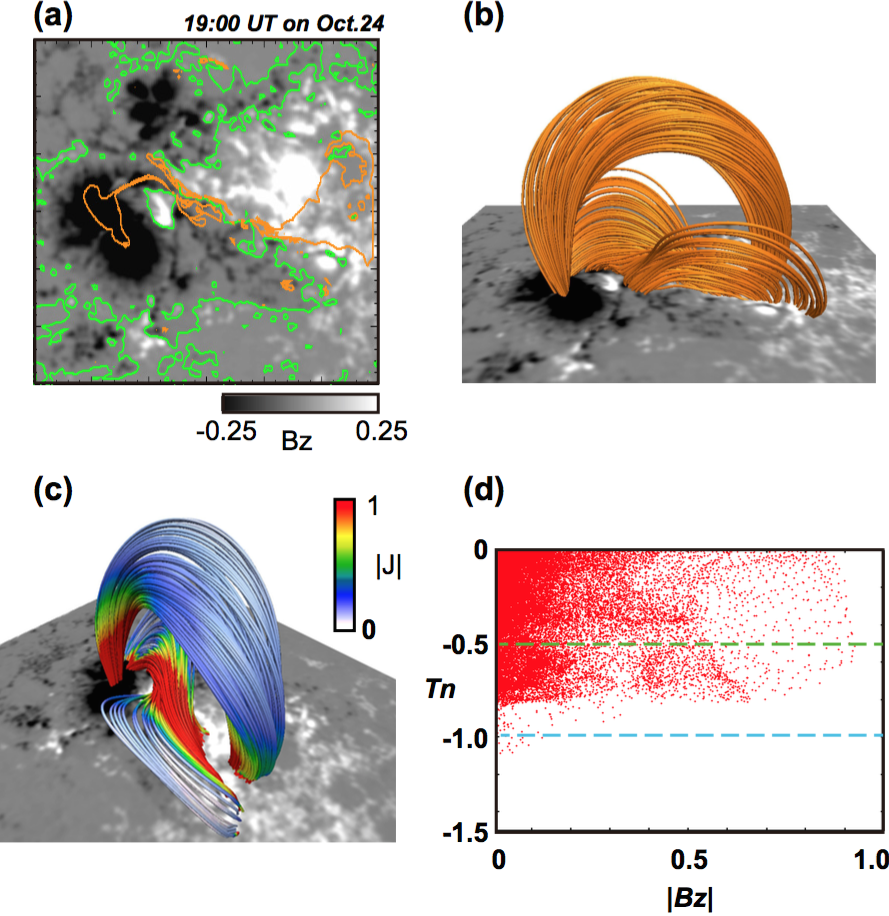}
  \caption{
       (a) Normal component $B_z$ of the magnetic fields taken  
            approximately 2h before the X3.1 class flare (19:00 UT), together 
            with contour of the half-turn twist (left-handed twist) calculated from 
            the NLFFF (orange lines)  and PIL (green lines). The calculation is 
            performed in the range of $0.2L_0 \leq x \leq 0.8L_0$ and
            $0.1L_0 \leq y \leq 0.7L_0$.
       (b) Three-dimensional structure of the field lines of the NLFFF (orange), 
            with $B_z$ distribution. All of the field lines are twisted with more 
            than a half-turn.
       (c) Same field lines as in (b) except that the color depends on the 
            strength of the current density $|\vec{J}|$.
       (d) Twist distribution versus $|B_z|$. Blue and green dashed 
            lines indicate one-turn and half-turn twist, respectively.
          }
  \label{f3}
  \end{figure}

   Figure \ref{f3}(a)  shows the contour of the half-turn twist obtained from the 
  NLFFF, mapped on the $B_z$ distribution approximately 2h before the X3.1-class 
  flare. Note that this twist is left-handed; therefore, its value of twist is 
  $T_n$=$-$0.5. The interior regions surrounded by the contours are occupied 
  by twisted lines with more than half-turn twist ("strongly twisted lines" hereafter); 
  {\it i.e.}, these regions are occupied by the footpoint of the flux tube. According to 
  this figure, the strongly twisted lines dominate in a wide region close to and 
  distant from the PIL. The former is along the PIL, whereas the later exits on the 
  magnetic flux constituting the major bipole field of the AR.  Figure \ref{f3}(b) shows 
  the 3D structure of the magnetic field lines traced from the regions inside of the 
  contours. We found flux tubes of different sizes and shapes;here, these are roughly 
  separated into two components. The first component is anchored in the major 
  bipole fields, and the second component is a small flux tube underneath the first. 
  Figure \ref{f3}(c) depicts the same field lines from the top with the strength of the 
  current density $|\vec{J}|$  superimposed.  According to this figure, the twisted lines  
  close to the solar surface have a strong current density, unlike that 
  accumulated in the large flux tube above them. Figure \ref{f3}(d) shows the twist 
  distribution versus $|B_z|$. We found that most of the field lines are less 
  than one turn, which implies that the magnetic field was stable against kink 
  mode instability before the X3.1-class flare. Therefore kink instability can be 
  eliminated as a candidate for the driver of the flare.
 
  \subsection{Magnetic Twist of the NLFFF After the X3.1 Class Flare}
\begin{figure}
  \epsscale{1.}
  \plotone{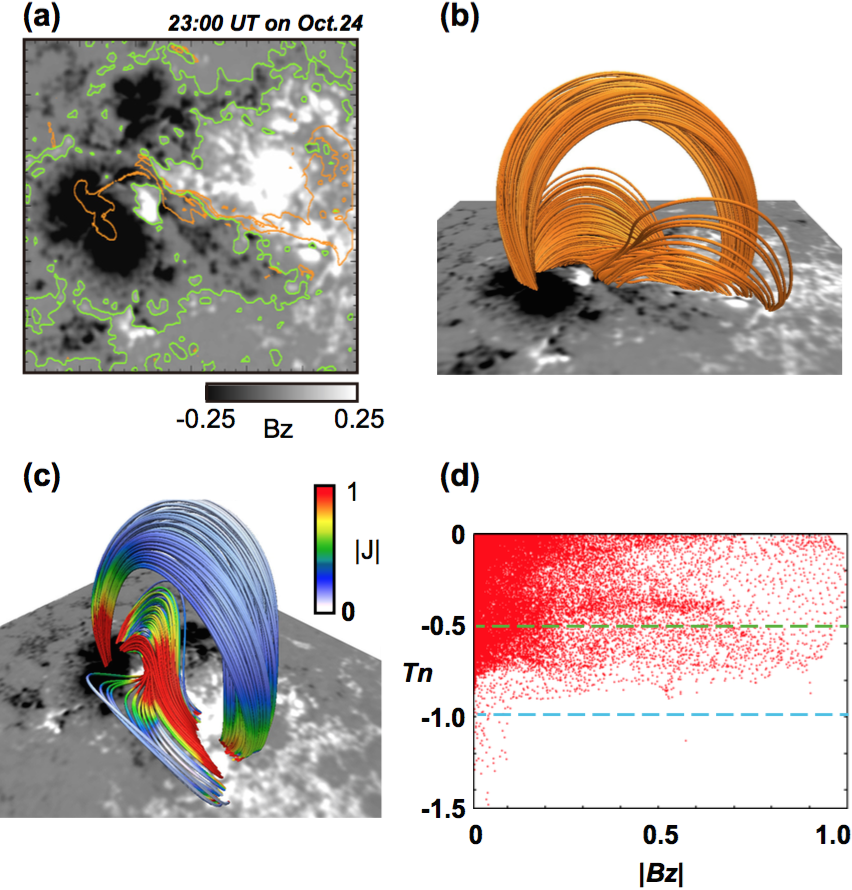}
  \caption{
       (a) Normal component $B_z$ of the magnetic fields taken at 23:00 UT 
            on October 24, which is approximately 2h after the X3.1-class flare, 
            together with contours of the half-turn twist calculated from the NLFFF 
            and PIL. The format is the same as that of  Figure \ref{f3}(a).        
      (b)  Field lines of the NLFFF with more than half-turn twist and the $B_z$ 
            distribution; the format is the same as that of Figure \ref{f3}(b).
      (c)  $|\vec{J}|$ is mapped on the field lines depicted in (b).
      (d)  Twist distribution versus $|B_z|$.  The format is the same as that of 
            Figure \ref{f3}(d).
           }
  \label{f4}
  \end{figure}

  Figure \ref{f4} shows the same plot as in Figure \ref{f3}, but for the second 
  instant (23:00 UT) which is approximately 2h after the flare. These profiles are 
  quite similar to those obtained before the flare, indicating that the current density 
  and magnetic twist accumulated before the flare seem to be conserved overall 
  (see Figures \ref{f4}(c) and (d)) despite the occurrence of the X-class flare. This 
  tendency is quite different from that found for other ARs, {e.g.}, AR10930 
  (\citealt{2012ApJ...760...17I}) or AR11158 (\citealt{2013ApJ...770...79I}).  Those 
  studies reported that most of the strongly twisted lines with more than half-turn 
  twist disappear after X-class flares, and some of them are converted into post-flare 
  loops.  This is a result of the process in which the magnetic twists accumulated 
  in the ARs are expelled to the upper corona. However, the results obtained from this 
  study are consistent with the fact that CMEs were not observed in via several X-class 
  flares.
  
    \subsection{Twisted Lines Producing the X3.1-class Flare}
\begin{figure}
  \epsscale{.8}
  \plotone{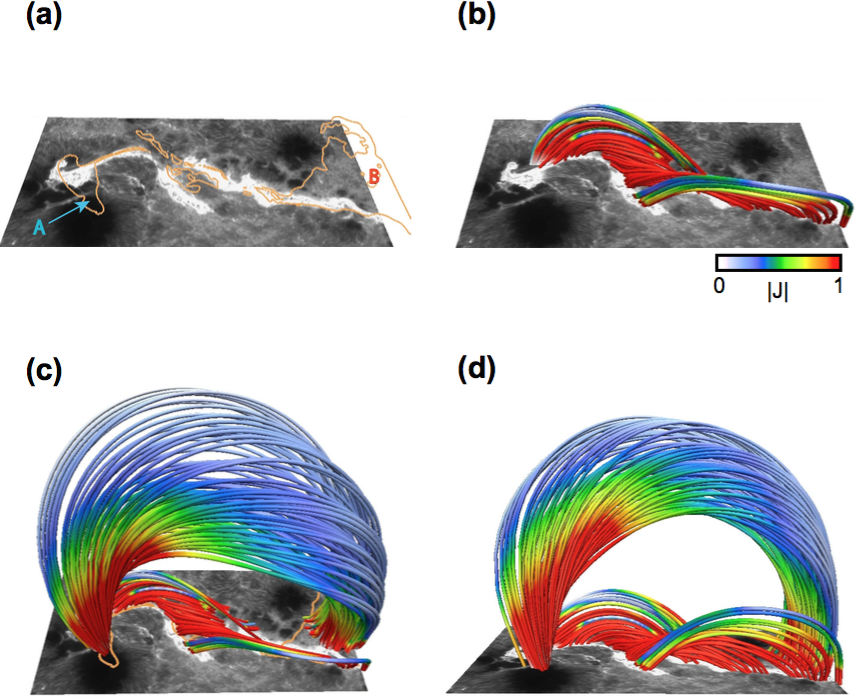}
  \caption{
           (a) Two-ribbon flares of X3.1-class flare observed in early phase of the 
                 flare by {\it Hinode}/Filtergram (FG) at 21:15 UT on October 24.
                 Orange lines indicate contours of half-turn twist ($T_n$=$-$0.5)
                 calculated from the NLFFF at19:00 UT, which is approximately 2h 
                 before the X3.1-class flare.
           (b)  NLFFF superimposed on (a) except without the twist contours. 
                 Field lines are selected both of whose footpoints are anchored 
                 on the enhanced region of {\ion{Ca}{2}}; color shows the strength 
                 of the current density $|\vec{J}|$. 
            (c) Strongly twisted lines with more than half-turn twist are added to 
                 the field lines shown in (b). Twisted lines start from region A (shown 
                 in (a)) and the other footpoint is anchored in region B (also shown 
                 in (a)).  Strong enhancement of {\ion{Ca}{2}} was not observed there 
                 during the flare. 
            (d) Side view of (c).
                 }
  \label{f5}
  \end{figure}

     In this section, we confirm which twisted lines produced the X3.1-class flare. 
    To do so, we compare the field lines of the NLFFF before the flare with the two-ribbon 
    flares obtained from {\ion{Ca}{2}} image taken from a filtergram of {\it Hinode}. 
    Figure \ref{f5}(a) shows the two-ribbon flares observed in the early phase of the 
    flare taken by {\it Hinode} together with the contours of half-turn twist obtained from 
    the NLFFF before the flare.  In Figure \ref{f5}(b) we further plot  selected field lines, 
    both of the footpoints of which are anchored in the enhanced region of {\ion{Ca}{2}}. 
    We found that these are sheared field lines, and a strong current is carried there. In 
    the tether-cutting model, sheared two-ribbon flares appear initially in regions 
    corresponding to the footpoints of sheared field lines in the onset phase owing to 
    reconnection between them. According to this scenario, these field lines close to the 
    solar surface would be related to the X3.1-class flare. 
    
     We add other twisted lines in Figure \ref{f5}(c) and show a side view in 
    Figure \ref{f5}(d);these lines are larger than the twisted lines shown in Figure \ref{f5}(b) 
    and straddle them. The twisted lines making up the large flux tube  are traced from region 
    A, marked in Figure \ref{f5}(a), and surrounded by a half turn twist contour, but strong 
    enhancement of {\ion{Ca}{2}}  was not observed during the flare. These field lines also 
    consist of strongly twisted lines with more than half-turn twist, but they are not related to 
    the X3.1-class flare. If these field lines are quite stable, they might suppress the 
    eruption accompanying with the flare.  We later discuss the stability of these twisted lines.
  
  \subsection{Location that Two-Ribbon Flares are Enhanced.}
    In the previous section, we found that in the first instant the twisted lines 
   were anchored in the flare-ribbon region, which might be a signature of 
   magnetic reconnection. On the other hand, all of the twisted lines are 
   unrelated to the X3.1-class flare; rather, the two-ribbon flare seemed to run 
   close to the edge of the region where the strongly twisted lines reside (see 
   Figure \ref{f5}(a)).
  
  Reconnection is considered to occur relativity easily in the  separatrix layer 
  (\citealt{2005LRSP....2....7L};
   \citealt{2006AdSpR..37.1269D};
   \citealt{2009ApJ...693.1029T})
  because it is a boundary between regions of magnetic field lines with different 
  connectivity, and formation of a thin current layer would be favorable there. 
  However, the observed two-ribbon flares appear in a limited area; {\it i.e.}, they 
  do not cover all of the separatrix on the solar surface. We clarify the structure 
  of the two-ribbon flare in comparison with the magnetic field by exploring the location 
  of the quasi-separatrix layer  (QSL) on the solar surface and the twist derived from the 
  NLFFF.
      
  We calculate the norm of the Jacobi matrix for magnetic field, following
  \cite{1996A&A...308..643D}: 
  \begin{equation}
  \displaystyle N(x,y) = sgn(B_z)\sqrt{ 
                         \sum_{i=1,2}
                         \left[ 
                           \left( \frac{\partial X_i}{\partial x} \right)^2 
                       +   \left( \frac{\partial X_i}{\partial y} \right)^2 
                         \right]
                              }, 
  \label{qsl}
  \end{equation} 
   where $(X_1, X_2)$ is the end point of a field line traced from the other footpoint 
  at position $(x,y)$ and $sgn(B_z)$ is defined by
  \[
  sgn(B_z) = \left\{
         \begin{array}{ll}
         -1 & \quad \mbox{$B_z<0$}, \\
          1 & \quad \mbox{$B_z>0$}.
         \end{array}\right.
  \] 
  The differentials in the x and y directions are approximated by 
  the grid interval. This means that the locations of the end points of two field lines 
  that are traced from these starting points across a large $N(x,y)$ value may differ 
  greatly.    
  
\begin{figure}
  \epsscale{1.}
  \plotone{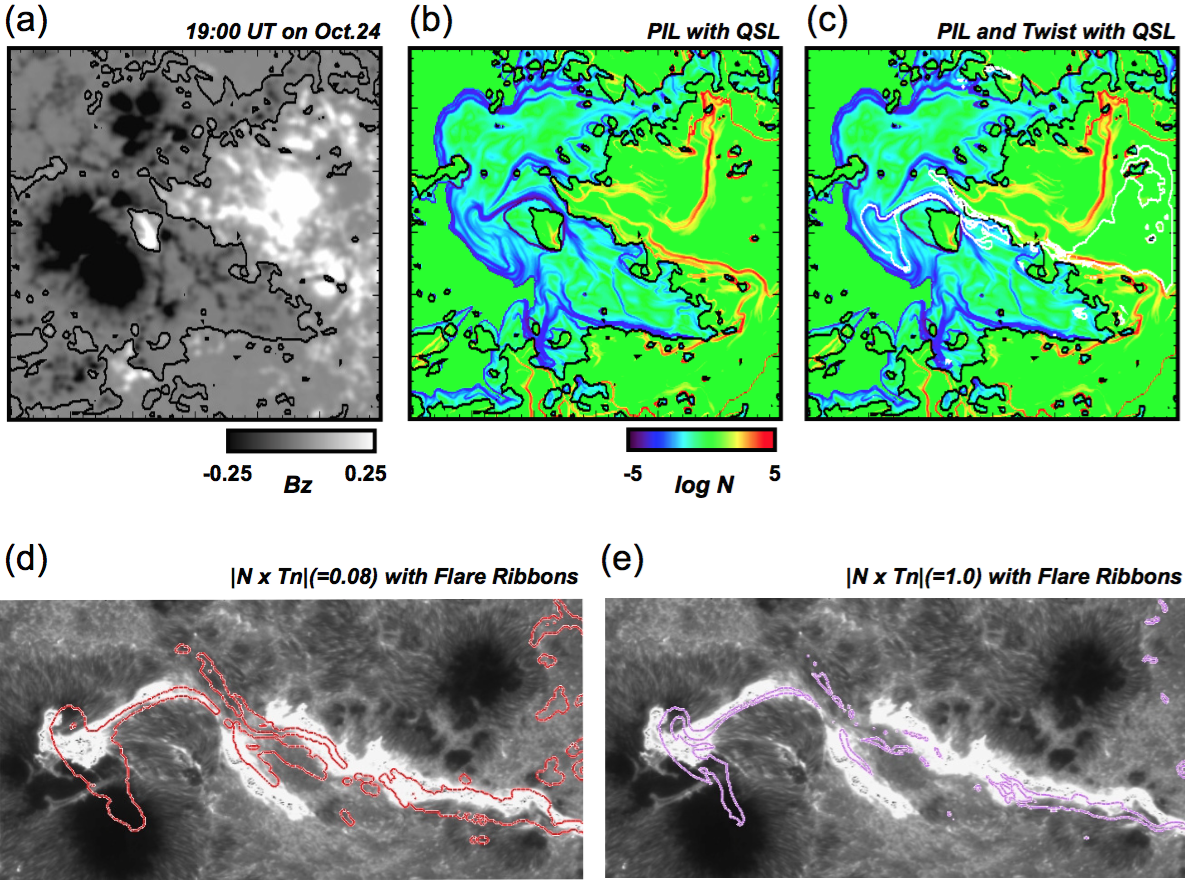}
  \caption{
       (a) $B_z$ distribution with PIL before the flare observed at 19:00 UT 
           on October 24.
       (b) Distribution of the norm $N(x,y)$ derived from the NLFFF 
             extrapolated from the photospheric field in (a) together with the 
             PIL, whose size corresponds to (a). 
       (c) PIL and twist contour $T_n$=$-$0.5 obtained from the NLFFF  
            plotted in black and white, respectively, with $N$ distribution.
       (d) Correspond of $|T_n \times N|$=0.08 (red lines) superimposed on 
             flare ribbons observed at 21:15 UT by {\it Hinode}.  The calculation 
             of $|T_n \times N|$ is satisfied for both $T_n<$-0.5 and $|N|>$1.  
       (e) Same as (d) except that the contour level of $|T_n \times N|$ 
             (purple) is 1.0.
          }
  \label{f6}
  \end{figure}

  Figures \ref{f6}(a) and (b) show the $B_z$ distribution with the PIL observed at 
  19:00 UT (before the flare) and a map of $N(x,y)$ (or more precisely, of its logarithm) 
  obtained from the NLFFF, respectively, both of which are plotted in the same area. 
  The connectivity of the field lines changes dramatically across the regions where 
  the $N(x,y)$ is large; consequently, some of  the QSLs would easily produce the 
  current layer, and reconnection might occur more easily there than in other areas. 
  On the other hand, because not all of the QSLs on the solar surface is 
  connected with solar flares, we need to extract the proper areas related to them 
  from the map of $N(x,y)$.  
  
   The magnetic twist is also an important proxy as well as $N$ in considering the 
  onset of solar eruptions, and it might help us to specify the {QSL} at which 
  the flare is strongly connected. Figure \ref{f6}(c) shows the twist distribution of 
  more than half-turn twist before the flare superimposed on Figure \ref{f6}(b). We 
  found that the boundaries of the twist can partially overlap with those of the strong 
  enhancement layers of $N(x,y)$, and these regions appear to be similar in shape 
  to the flare ribbons shown in Figure \ref{f5}(a). To confirm this, we further calculate 
  $|T_n \times N|$ in a range of $T_n \leq -0.5$ and $|N| \geq 1$, and plotted it in 
  Figure \ref{f6}(d) with the flare ribbons. The contours of $|T_n \times N|$ roughly 
  capture the distribution of the flare ribbons. Furthermore, in Figure \ref{f6}(e), we 
  plot the $|T_n \times N|$ with another contour level having a higher value than that 
  shown in Figure \ref{f6}(d). We found that most of the contours are along the edge 
  of the ribbons; {\it i.e.,} the strongly enhanced layer of $N$ would correspond to the 
  boundary of the flare ribbons. This interpretation is consistent with a growth 
  of  the flare ribbons being inhibited by the QSL, as suggested by 
  \cite{2011arXiv1109.0381C}. These results suggest that the flare ribbons are not 
  related to just the intersection of the QSL with the solar surface or the twistedness,  
  but that both pieces of  information are needed.  From another view point, interestingly, 
  the shape of the flare ribbons might be predicted from the NLFFF before a flare.
   
  \section{Discussion}
  \subsection{Stability Against the Torus Instability}
  In the previous section, we first saw that the NLFFF reconstructed 2h 
  before the X3.1-class flares is stable to kink mode instability. We further 
  discuss the stability against torus instability as well. The torus instability 
  is caused by a process, in which the force balance acting on the flux tube 
  is broken, where the upward force is the hoop force due to a current in the 
  flux tube and the downward force is contributed by the surroundings. A 
  criterion where the instability can occur is defined by the decay 
  index (\citealt{2006PhRvL..96y5002K}),
  \begin{equation}
    n = -\frac{z}{|\vec{B}|}\frac{\partial |\vec{B}|}{\partial z},
  \label{di}
  \end{equation}
  which is calculated from the external field surrounding the flux tube. 
  Several earlier studies noted that instability happens at  the condition 
  $n$=1.5 ({\it e.g.}, \citealt{2007AN....328..743T}).

\begin{figure}
  \epsscale{1.}
  \plotone{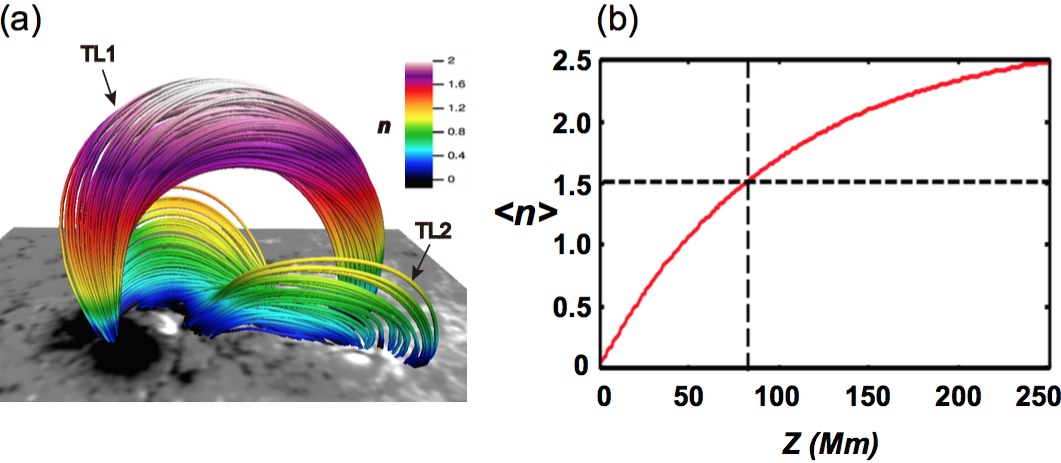}
  \caption{
       (a) Three-dimensional magnetic field lines of the NLFFF with more 
             than half-turn twist are plotted. Color indicates the value of the decay 
             index $n$, and the $B_z$ distribution on the bottom surface, all of 
             which are observed at 19:00 UT (before the flare). Each arrow  
             indicates a flux tube reconstructed in the NLFFF.
       (b)  Average decay index defined by Equation (\ref{ave_di}) in height 
             profile, which is calculated in a range of $0.25L_0 \leq x \leq 0.75L_0$ 
             and $0.2L_0 \leq y \leq 0.7L_0$ for $N_x$=$N_y$=150. 
             Vertical and horizontal dashed lines cross at the threshold of the torus 
             instability inferred from the theory. 
          }
  \label{f7}
  \end{figure}

   Figure \ref{f7}(a) shows the twisted field lines with more than a half-turn 
  twist at 19:00 UT, {\it i.e.},  approximately 2h before the X3.1-class flare, 
  where the colors represent the different values of $n$. We clearly found that the 
  upper half of TL1 indicated in Figure \ref{f7}(a) satisfies $n \geq1.5$, even 
  though most of TL2 meets $n \leq1.0$. Figure \ref{f7}(b) shows the height 
  profile of the decay index  where this decay index is averaged in the $x-y$ 
  plane 
  \begin{equation}
  <n(z)> = \frac{1}{N_xN_y}\sum_{i,j}^{N_x, N_y} n(x_i,y_j,z), 
  \label{ave_di}
  \end{equation}
  where $N_x$ and $N_y$ are grid number in the calculated area in the $x$ 
  and $y$ directions, respectively. According to these results, $n$=1.5 
  corresponds to approximately 70-80(Mm). This result is consistent with 
  \cite{2015ApJ...804L..28S}, which further reported that this height is greater 
  than that estimated in other ARs ({\it e.g.,} AR11158, AR11429). From our 
  results, we found that most of the twisted lines connected with the X-class 
  flares, {\it i.e.}, TL2 in Figure \ref{f7}(a), reside in an area that is stable against 
  the torus instability. On the other hand, the top part of TL1 exceeds even 
  $n$=2.0 in its twistness. The twisted lines obviously meet the unstable condition; 
  nevertheless, CMEs associated with the huge flare were not observed from this 
  AR.

\begin{figure}
  \epsscale{1.}
  \plotone{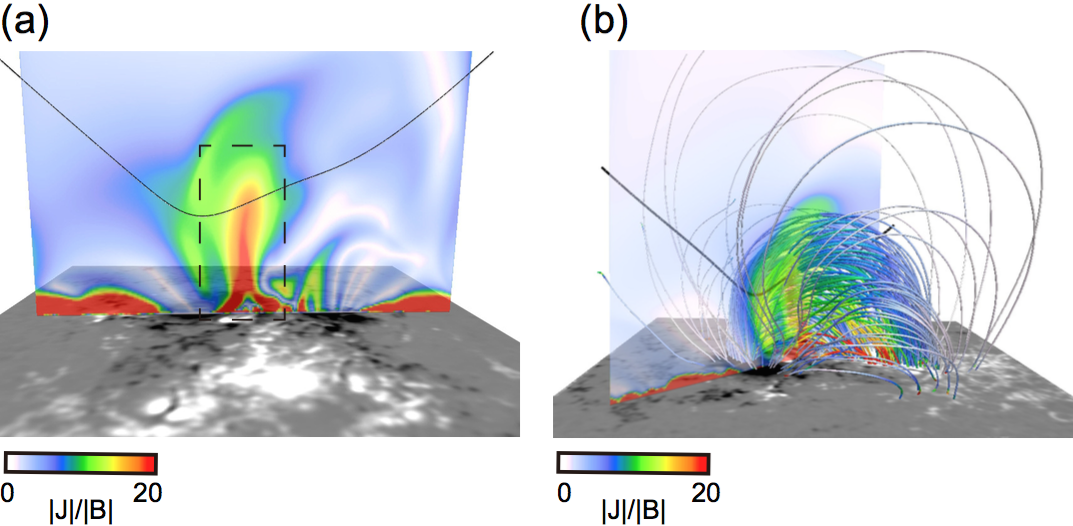}
  \caption{
       (a) $|\vec{J}|/|\vec{B}|$ drawn on a vertical cross section, 
            that is a side view from the west side obtained from the NLFFF before 
            the flare. Black solid line indicates contour of $n$=1.5, and vertically 
            elongated current is surrounded by the dashed square.
       (b) Field lines colored according to $|\vec{J}|/|\vec{B}|$ superimposed 
            on the $B_z$ distribution. Format of the vertical cross section and 
            solid line are identical to those in (a) except that the angle is different. 
          }
  \label{f8}
  \end{figure}

   The major reason for the suppressing of the instability of the upper flux tube, 
  TL1, is that the electric current flowing in the flux tube is not well approximated 
  by a single line current loop, which was used in the theory of torus instability 
  (\citealt{2006PhRvL..96y5002K}). The decay index is derived from the external field, 
  in which the flux tube approximated by the loop of a single current line stays in an 
  equilibrium state, but there is no guarantee that this scenario applies in this magnetic 
  field in which the current is widely distributed as shown in Figure \ref{f8}(a), where the 
  current is divided by $|\vec{B}|$ at each place. Figure \ref{f8}(b) shows the field 
  lines colored according to $|\vec{J}|/|\vec{B}|$. The vertical extended current produces 
  a long vertical flux tube, which deviates greatly from the flux tube derived from a single 
  loop current. From another viewpoint, because variable magnetic flux tubes of different 
  size and shapes cross the vertically extended electric current distribution, the stability of 
  these complex magnetic structure may not be easily determined. For instance,  we 
  consider a simple situation in which a double-decker flux tube (\citealt{2012ApJ...756...59L}; 
  \citealt{2014ApJ...789...93C}) is assumed and each flux tube has a line current in the 
  same direction. Then the downward force, {\it i.e.}, the pinch force, due to the current in 
  the lower twisted lines acts on the upper twisted lines. Obviously this pinch force causes 
  the upper twisted lines to enter a  stable condition. In other words, because the hoop 
  force on the upper flux tube is weakened, we can suggest that the height of the typical 
  loop taken up in the instability would decrease. Nevertheless, the hoop force works 
  efficiently on the lower twisted lines (TL2), but these are mostly stable; {\it i.e.}, this 
  mechanical interaction between the different twisted flux ropes (TL1 and TL2) might 
  keep the stability. 

    \subsection{MHD Simulation to Confirm the Stability}
  In this paper, we discussed the stability of the NLFFF, using the magnetic twist as expressed
  in Equation  (\ref{twist}) together with decay index as defined by Equation (\ref{di}).  However, 
  this analysis is not rigorous because these criteria were derived under certain assumptions 
  and there is no guarantee these assumptions are also valid for our NLFFF model. Therefore, in order 
  to obtain further confirmation of the stability of the configuration, we performed an MHD 
  simulation using the NLFFF at 21:00 UT as the initial condition, and neglecting external forces. 
  On the other hand, since the NLFFF is calculated numerically, it is not completely in force 
  balance and hence a residual force exists, which plays the role of a perturbation. Therefore if the 
  NLFFF would be unstable, the unstable mode should grow exponentially, resulting in an increase 
  of the kinetic energy and a deformation of the magnetic structure.  
  
  The MHD simulation is identical to that performed in \cite{2014ApJ...788..182I} and  
  \cite{2015ApJ...788..182I}, whose Equations (\ref{rho_eq})-(\ref{div_eq}) are being solved but 
  here a constant resistivity $\eta$=$5.0\times 10^{5}$ is chosen. The velocity is controlled only 
  on the nearby side boundaries according to the manner described in  Section 2.2, so that the 
  numerical noise could be suppressed. Regarding to the boundary condition, the normal component of the 
  magnetic field $B_n$ is fixed on all boundaries, and the horizontal components $\vec{B_t}$ 
  are determined by the induction equation. The velocity field is fixed to zero and the same  
  von Neumann boundary condition as in the NLFFF calculation is imposed on $\phi$.
     
\begin{figure}
  \epsscale{1.}
  \plotone{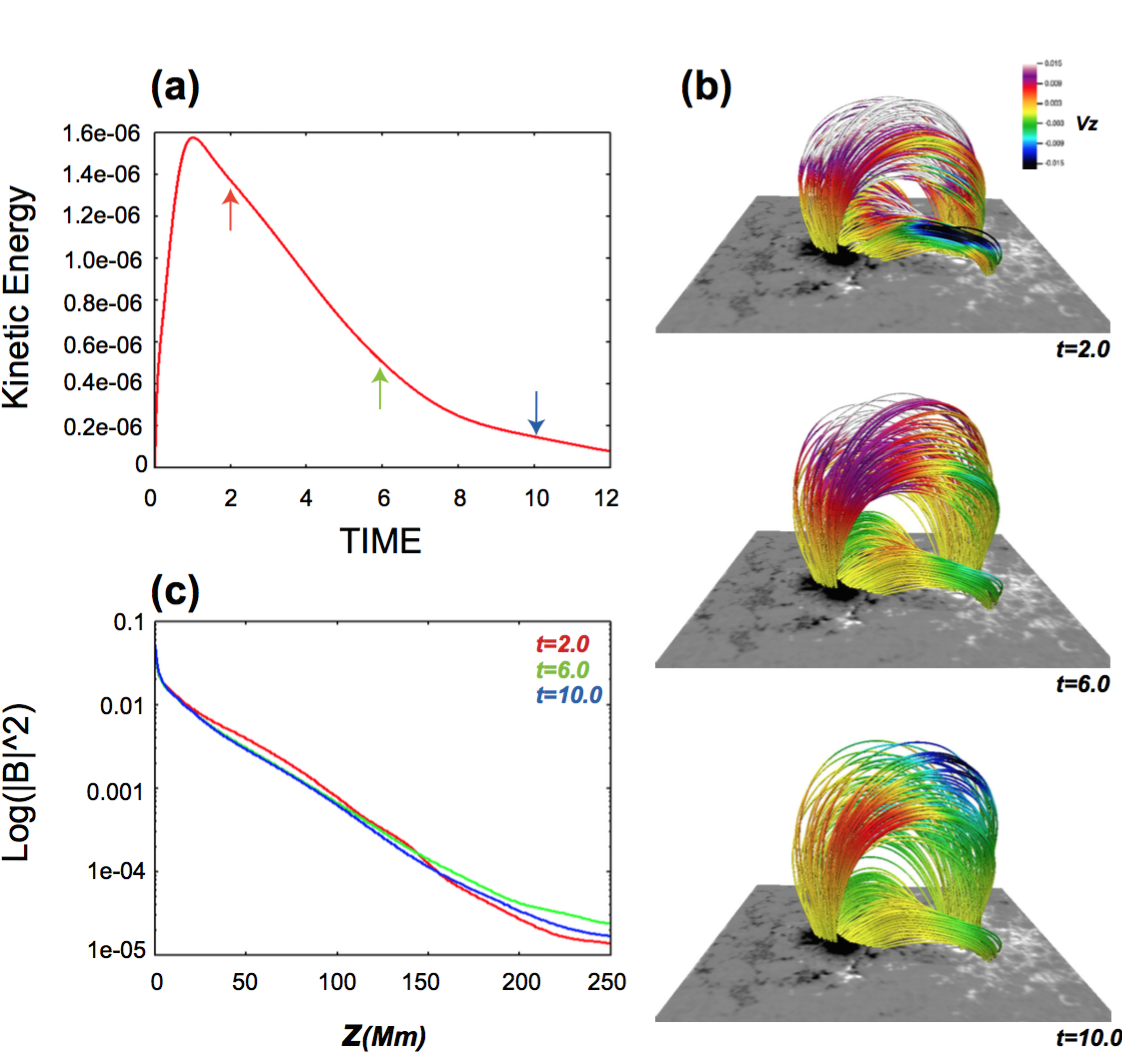}
  \caption{
                Results from the MHD simulation.
                (a) The temporal evolution of the kinetic energy, that is marked with 
                     each colored arrow at $t$=2.0, $t$=6.0, and $t$=10.0, respectively.  
                (b) The temporal evolution of the 3D magnetic structures at each time 
                     marked with the colored arrows in (a) where the color on the field 
                     lines indicates a distribution of the vertical velocity.
               (c)  Vertical profiles of $|\vec{B}|^2$ at center of the numerical box 
                     where the red, green and blue lines correspond to those at $t$=2.0, 
                     $t$=6.0, and $t$=10.0, respectively.
          }
  \label{f9}
  \end{figure}

   Figure \ref{f9}(a) shows the temporal evolution of the kinetic energy 
  ($E_{k}=\int \rho |\vec{v}^2|$/2 dV). Although the value once increases suddenly around $t$=1.0, 
  the growth stops there after and quickly decreases when $t>$1.0. This sudden increase is excited 
  by the residual velocity which cannot be removed from the NLFFF. Figure \ref{f9}(b) exhibits the 
  temporal evolution of the 3D magnetic structures. The  drawn magnetic field lines exhibit little 
  changes in structures over a long span of time, as seen at $t$=2.0, $t$=6.0 and $t$=10.0. We found 
  that the vertical velocity on the twisted lines is also getting small in the temporal evolution.  Finally, 
  we also confirm the facts that whether or not the boundary condition prevents the growth of the 
  instability. Figure \ref{f9}(c) shows the temporal evolution of the vertical profile on $|\vec{B}|^2$ 
  where $|\vec{B}|^2=B_x^2+B_y^2+B_z^2$. We can find an imperceptible variation around the top 
  boundary, {\it i.e.,} no enhancement in the vicinity of the top boundary would give no feedback to 
  the flux tube. Therefore, these results also support that the NLFFF is stable against the MHD 
  instability.  
  
  \section{Summary}
  This paper presents the structure and stability analysis of the 3D magnetic 
  fields of large solar AR 12192 reconstructed under the NLFFF approximation 
  before and after the X3.1-class flare that occurred around 21:00 UT on October 24 
  2014. This AR contains two huge major bipoles and also a complex magnetic 
  structure of mixed positive and negative polarities close to the PIL. The 3D configuration 
  of the magnetic field is also not so simple. We found a multiple-flux-tube system 
  composed of variable twisted lines of different shapes and sizes. This includes a large 
  flux tube anchored to the main polarities, under which several flux tubes exist along 
  the PIL. In particular, the later configuration(lower one) is favorable for the tether-cutting 
  reconnection, and the location of their footpoints corresponds to the area in which 
  two-ribbon flares are strongly enhanced during the X3.1-class flare. This means that the 
  upper twisted lines are not related to that flare.
  
  The stability of the magnetic configuration was also analyzed. We estimated the 
  magnetic twist to confirm the stability against the kink instability; since the most 
  of the twisted lines have a twist less than one turn, we, therefore, conclude that the 
  NLFFF is in stable state against kink instability. On the other hand, some of the upper 
  twisted lines are in the region of over the critical decay index for the torus instability, 
  even though an eruption was not observed from this AR in the middle of October 2014. 
  Hence we suggest that this estimation of the critical decay index cannot be applied 
  to the multiple-flux-tube system. The existing theory was established where 
  current was confined to a single loop current inside the flux tube, but in practice the 
  current is distributed in the solar corona. For the present study, because we have to 
  consider the interaction of each flux tube, it would not be easy to estimate the loop height 
  at which the instability is excited. Although all of these stability analyses are performed 
  based on the linear stability analysis,  we further numerically confirm the stability through 
  MHD simulation.
  
   Finally we discussed the location at which the two ribbon flares appear. On the 
  basis of our results, we suggested that the location at which flare ribbons appear 
  might be specified using both QSL defined by large $N$ and a strong magnetic 
  twist,  in particular, a region in which strong enhancement of $N$ corresponds
  to the boundaries of the flare ribbons. This simply means that the discontinuity 
  region composed of strongly twisted lines, especially those with more than half-turn twist, 
  would be a candidate for producing huge solar flares. With future development, we expect 
  that this concept may be a useful method for predicting solar flares which in turn may lead 
  to the space weather prediction.

  \acknowledgments
   We are grateful to referee for his/her constructive comments and Dr.\ Vinay Shanker Pandey 
   for reading the proof version of this paper. This work was supported by JSPS KAKENHI Grant 
   Numbers 23340045, and 15H05814. S.\ I.\ sincerely thanks to the Alexander von Humboldt  
   foundation for supporting his stay in Germany as well as this work. The computational work 
   was carried out within the computational joint research program at the Solar-Terrestrial Environment 
   Laboratory, Nagoya University. Computer simulation was performed on the Fujitsu PRIMERGY CX400 
   system of the Information Technology Center, Nagoya  University. The computational resources 
   for the data analysis and visualization system by using resource of the OneSpaceNet in 
   the NICT Science Cloud and VAPOR
   (\citealt{2005SPIE.5669..284C},
   \citealt{2007NJPh....9..301C}).

\clearpage

  \clearpage

  \clearpage

  \clearpage
  
  \clearpage
  
  \clearpage

  \clearpage

  \clearpage

  \clearpage
  
  \clearpage





\begin{thebibliography}{}
\bibitem[Amari et al.(1996)]{1996ApJ...466L..39A} Amari, T., Luciani, J.~F., Aly, J.~J., 
           \& Tagger, M.\ 1996, \apjl, 466, L39 

\bibitem[Amari et al.(2014)]{2014Natur.514..465A} Amari, T., Canou, A., \& 
         Aly, J.-J.\ 2014, \nat, 514, 465 

\bibitem[Berger \& Prior(2006)]{2006JPhA...39.8321B} Berger, M.~A., \& 
         Prior, C.\ 2006, Journal of Physics A Mathematical General, 39, 8321 

\bibitem[Bobra et al.(2014)]{2014SoPh..289.3549B} Bobra, M.~G., Sun, X., 
         Hoeksema, J.~T., et al.\ 2014, \solphys, 289, 3549
                  
\bibitem[Borrero et al.(2011)]{2011SoPh..273..267B} Borrero, J.~M., 
         Tomczyk, S., Kubo, M., et al.\ 2011, \solphys, 273, 267

\bibitem[Canou et al.(2009)]{2009ApJ...693L..27C} Canou, A., Amari, T., 
         Bommier, V., et al.\ 2009, \apjl, 693, L27 

\bibitem[Chen et al.(2011)]{2011arXiv1109.0381C} Chen, P.~F., Su, J.~T., Guo, Y., 
            \& Deng, Y.~Y.\ 2011, arXiv:1109.0381 

\bibitem[Cheng et al.(2014)]{2014ApJ...789...93C} Cheng, X., Ding, M.~D., 
               Zhang, J., et al.\ 2014, \apj, 789, 93 
         
\bibitem[Clyne \& Rast(2005)]{2005SPIE.5669..284C} Clyne, J., \& Rast, M.\ 
         2005, \procspie, 5669, 284 

\bibitem[Clyne et al.(2007)]{2007NJPh....9..301C} Clyne, J., Mininni, P., 
         Norton, A., \& Rast, M.\ 2007, New Journal of Physics, 9, 301 

\bibitem[Dedner et al.(2002)]{2002JCoPh.175..645D} Dedner, A., Kemm, F., 
         Kr{\"o}ner, D., et al.\ 2002, Journal of Computational Physics, 
         175, 645 

\bibitem[D{\'e}moulin \& Aulanier(2010)]{2010ApJ...718.1388D} D{\'e}moulin, P., 
           \& Aulanier, G.\ 2010, \apj, 718, 1388 

\bibitem[Demoulin et al.(1996)]{1996A&A...308..643D} Demoulin, P., 
             Henoux, J.~C., Priest, E.~R., \& Mandrini, C.~H.\ 1996, \aap, 308, 
             643 
         
\bibitem[D{\'e}moulin(2006)]{2006AdSpR..37.1269D} D{\'e}moulin, P.\ 2006, 
             Advances in Space Research, 37, 1269         
              
         
\bibitem[Guo et al.(2010)]{2010ApJ...725L..38G} Guo, Y., Ding, M.~D., 
         Schmieder, B., et al.\ 2010, \apjl, 725, L38

\bibitem[Hoeksema et al.(2014)]{2014SoPh..289.3483H} Hoeksema, J.~T., Liu, Y.,
         Hayashi, K., et al.\ 2014, \solphys, 289, 3483

\bibitem[Inoue et al.(2014a)]{2014ApJ...780..101I} Inoue, S., Magara, T., 
               Pandey, V.~S., et al.\ 2014a, \apj, 780, 101
          
\bibitem[Inoue et al.(2014b)]{2014ApJ...788..182I} Inoue, S., Hayashi, K., 
         Magara, T., Choe, G.~S., \& Park, Y.~D.\ 2014b, \apj, 788, 182 
         
\bibitem[Inoue et al.(2015)]{2015ApJ...788..182I} Inoue, S., Hayashi, K., 
         Magara, T., Choe, G.~S., \& Park, Y.~D.\ 2015, \apj, 788, 182 

\bibitem[Inoue et al.(2011)]{2011ApJ...738..161I} Inoue, S., Kusano, K., 
         Magara, T., Shiota, D., \& Yamamoto, T.~T.\ 2011, \apj, 738, 161
         
 \bibitem[Inoue et al.(2012)]{2012ApJ...760...17I} Inoue, S., Shiota, D., 
         Yamamoto, T.~T., et al.\ 2012, \apj, 760, 17
         
 \bibitem[Inoue et al.(2013)]{2013ApJ...770...79I} Inoue, S., Hayashi, K., 
         Shiota, D., Magara, T., \& Choe, G.~S.\ 2013, \apj, 770, 79 
             
 \bibitem[Ji et al.(2003)]{2003ApJ...595L.135J} Ji, H., Wang, H., Schmahl, 
          E.~J., Moon, Y.-J., \& Jiang, Y.\ 2003, \apjl, 595, L135

 \bibitem[Jiang et al.(2014)]{2014ApJ...780...55J} Jiang, C., Wu, S.~T., 
             Feng, X., \& Hu, Q.\ 2014, \apj, 780, 55 

 \bibitem[Joshi et al.(2014)]{2014ApJ...795....4J} Joshi, N.~C., Magara, T., 
         \& Inoue, S.\ 2014, \apj, 795, 4 

 \bibitem[Kataoka et al.(2014)]{2014SpWea..12..380K} Kataoka, R., Sato, T., 
          Kubo, Y., et al.\ 2014, Space Weather, 12, 380
    
 \bibitem[Kliem \& T{\"o}r{\"o}k (2006)]{2006PhRvL..96y5002K} Kliem, B., 
         T{\"o}r{\"o}k, T. \ 2006, Physical Review Letters, 96, 255002   
    
 \bibitem[Kosugi et al.(2007)]{2007SoPh..243....3K} Kosugi, T., Matsuzaki, K.,
          Sakao, T., et al.\ 2007, \solphys, 243, 3  
         
 \bibitem[Leka et al.(2009)]{2009SoPh..260...83L} Leka, K.~D., Barnes, G., 
          Crouch, A.~D., et al.\ 2009, \solphys, 260, 83 
          
 \bibitem[Lemen et al.(2012)]{2012SoPh..275...17L} Lemen, J.~R., Title, A.~M., 
         Akin, D.~J., et al.\ 2012, \solphys, 275, 17
                           
 \bibitem[Liu et al.(2012)]{2012ApJ...756...59L} Liu, R., Kliem, B., 
          T{\"o}r{\"o}k, T., et al.\ 2012, \apj, 756, 59

  \bibitem[Longcope(2005)]{2005LRSP....2....7L} Longcope, D.~W.\ 2005, Living 
  Reviews in Solar Physics, 2, 7 

 \bibitem[Metcalf(1994)]{1994SoPh..155..235M} Metcalf, T.~R.\ 1994, 
         \solphys, 155, 235 
        
 \bibitem[Moore et al.(2001)]{2001ApJ...552..833M} Moore, R.~L., 
          Sterling, A.~C., Hudson, H.~S., \& Lemen, J.~R.\ 2001, \apj, 552, 
          833 
            
 \bibitem[Savcheva et al.(2015)]{2015arXiv150603452S} Savcheva, A., 
           Pariat, E., McKillop, S., et al.\ 2015, arXiv:1506.03452         
            
 \bibitem[Scherrer et al.(2012)]{2012SoPh..275..207S} Scherrer, P.~H., 
          Schou, J., Bush, R.~I., et al.\ 2012, \solphys, 275, 207 

 \bibitem[Sun et al.(2015)]{2015ApJ...804L..28S} Sun, X., Bobra, M.~G., 
          Hoeksema, J.~T., et al.\ 2015, \apjl, 804, L28 

\bibitem[Titov et al.(2009)]{2009ApJ...693.1029T} Titov, V.~S., Forbes, T.~G., 
              Priest, E.~R., Miki{\'c}, Z., \& Linker, J.~A.\ 2009, \apj, 693, 1029

 \bibitem[T{\"o}r{\"o}k \& Kliem(2007)]{2007AN....328..743T} T{\"o}r{\"o}k, T.,
          \& Kliem, B.\ 2007, Astronomische Nachrichten, 328, 743 

 \bibitem[Tsuneta et al.(2008)]{2008SoPh..249..167T} Tsuneta, S., Ichimoto, K.,
          Katsukawa, Y., et al.\ 2008, \solphys, 249, 167     

 \bibitem[Wiegelmann \& Sakurai(2012)]{2012LRSP....9....5W} Wiegelmann, T., \&
          Sakurai, T.\ 2012, Living Reviews in Solar Physics, 9, 5 

 \bibitem[Wiegelmann et al.(2006)]{2006SoPh..233..215W} Wiegelmann, T., 
          Inhester, B., \& Sakurai, T.\ 2006, \solphys, 233, 215 
         
 \bibitem[Yashiro et al.(2006)]{2006ApJ...650L.143Y} Yashiro, S., Akiyama, S., 
          Gopalswamy, N., \& Howard, R.~A.\ 2006, \apjl, 650, L143

 \bibitem[Zuccarello et al.(2014)]{2014ApJ...785...88Z} Zuccarello, F.~P., 
          Seaton, D.~B., Mierla, M., et al.\ 2014, \apj, 785, 88

 \end{thebibliography}
\end{document}